\documentstyle[12pt,titlepage]{article}
\begin{document}
\begin{titlepage}
\parindent 0pt
{\Large\bf
Absence of re-entrant phase transition
of the antiferromagnetic Ising model on the simple cubic lattice:
Monte Carlo study of the hard-sphere lattice gas
}

\bigskip

{\large\bf Atsushi Yamagata}

{\it
Department of Physics, Tokyo Institute of Technology,
Oh-okayama, Meguro-ku, Tokyo 152, Japan
}

\bigskip

\begin{description}
\item[Running title]
Absence of re-entrant phase transition of
antiferromagnetic Ising model

\item[Keywords]
Antiferromagnetic Ising model, Hard-sphere lattice gas,
Monte Carlo method

\item[PACS classification codes]
02.70.Lq, 05.50.+q, 64.60.Cn, 75.10.Hk
\end{description}

\bigskip

{\bf Abstract}

We perform the Monte Carlo simulations of the hard-sphere lattice gas
on the simple cubic lattice with nearest neighbour exclusion.
The critical activity is estimated, $z_{\rm c} = 1.0588 \pm 0.0003$.
Using a relation between the hard-sphere lattice gas and
the antiferromagnetic Ising model in an external magnetic field,
we conclude that there is no re-entrant phase transition of
the latter on the simple cubic lattice.
\end{titlepage}

\section{Introduction}
The antiferromagnetic Ising model shows a phase transition
in an external magnetic field but the ferromagnetic one has
a critical point only in zero field.
The critical line surrounds the antiferromagnetic ordered phase
\cite{SauerTemperley40,Garrett51exp}.
A Hamiltonian is
\[
{\cal H}
=
|J| \sum_{\langle i j \rangle} s_{i}\,s_{j} -H \sum_{i} s_{i},
\]
where $s_{i}$ is an Ising spin variable located {\em i\/}th lattice
site and which takes on the value $+1$ and $-1$.
The first summation is over all nearest neighbour pairs on a lattice
and the second over all lattice sites.
$J(<0)$ is the exchange interaction.
$H$ is an external magnetic field.
Many authors have studied the system by various methods:
Bethe approximation
\cite{Ziman51},
mean field approximation
\cite{Garrett51},
constant coupling approximation
\cite{KasteleijnVanKranendonk56},
Kikuchi approximation
\cite{Burley61},
series expansions
\cite{Bienenstock66,BienenstockLewis67},
Monte Carlo simulations
\cite{Shirley72,Metcalf73,Shirley77,Landau77,Rapaport78,
MetcalfYang78},
transfer matrix
\cite{MetcalfYang78},
renormalization group with periodic cell clusters
\cite{SubbaswamyMahan76,MahanClaro77}
phenomenological renormalization group and transfer matrix
\cite{Sneddon79,Racz80},
exact calculations of an interfacial model
\cite{MullerHartmannZittartz77},
finite-size scaling and transfer matrix
\cite{BloteWu90},
analyses of a 16-vertex model
\cite{WuWu90}.

On a bipertite lattice there is a critical field,
$H_{\rm c} = q |J|$ where $q$ is a coordination number.
At $T=0$ all of the nearest neighbour pairs is antiparallel for
$|H|<H_{\rm c}$ and is parallel for $|H|>H_{\rm c}$.
At $H=0$ the antiferromagnetic Ising model is equivalent to
the ferromagnetic one.
Then in the former a phase transition occurs
at the N\'eel temperature, $T_{\rm N}$,
which is equal to the critical temperature of the latter.
The three points, $(T,H)$ = $(0,H_{\rm c})$, $(T_{\rm N},0)$,
and $(0,-H_{\rm c})$, are on the critical line.
Since the phase diagram is symmetrical for $H=0$,
we consider non-negative $H$ case.

When the slope of the critical line at $(T,H) = (0,H_{\rm c})$
is positive,
a re-entrant phase transition occurs.
The system is in the paramagnetic, the antiferromagnetic ordered,
and
the paramagnetic phase as $T$ is decreased
when $H$ is slightly above $H_{\rm c}$.
It is negative on the square lattice
\cite{Racz80,BloteWu90,WuWu90}
and is positive on the body-centred cubic lattice
\cite{Shirley72,Shirley77,Landau77,Racz80}.
For the simple cubic lattice there is no conclusive result
\cite{MahanClaro77,Racz80}.
The purpose of this paper is to identify
whether there is the re-entrant phase transition
of the antiferromagnetic Ising model
on the simple cubic lattice or not.

We carry out Monte Carlo simulations of a hard-sphere lattice gas
rather than the antiferromagnetic Ising model.
In the next section we review a relation between them.
We describe a Monte Carlo algorithm for the hard-sphere lattice gas
and define physical quantities measured in section~\ref{sec:mcs}.
In section~\ref{sec:mcr} we present Monte Carlo results.
A summary is given in section~\ref{sec:sum}.

\section{Hard-sphere lattice gas}
The slope of the critical line relate to the critical activity
of a hard-sphere lattice gas with nearest neighbour exclusion.
The atoms can occupy lattice sites under two conditions:
only one atom may occupy a lattice site and
they interact with infinite repulsion of nearest neighbour pairs.
There are many studies of the system:
series expansions
\cite{Domb58,Temperley59,Burley60,GauntFisher65,Gaunt67},
finite-size scaling and transfer matrix
\cite{KamieniarzBlote93},
Bethe and ring approximations
\cite{Burley61b},
transfer matrix
\cite{Runnels65,RunnelsCombs66,ReeChesnut66},
corner transfer matrix and series expansions
\cite{Baxteretal80},
exact calculations
\cite{Baxter80},
Monte Carlo simulations
\cite{Meirovitch83}.

The grand partition function is
\begin{equation}
{\mit \Xi_{V}(z)}
=
\sum_{N} z^{N}\,Z_{V}(N),
\label{eqn:gpf}
\end{equation}
where $z$ is an activity and $V$ is the number of lattice sites.
$Z_{V}(N)$ is the number of configurations in which
there are $N$ atoms in a lattice of $V$ sites.
At $z=+\infty$ a ground state configuration is
that the atoms occupy all the sites of
one sublattice and the other is vacant.
There is no atom at $z=0$.
The critical activity, $z_{\rm c}$, relates to the slope,
$a^{\ast}$, of the critical line at $(T,H)=(0,H_{\rm c})$
defined by $H=H_{\rm c}+a^{\ast} k_{\rm B} T$
\cite{Racz80,Temperley59,Burley60,Baxteretal80}:
\begin{equation}
a^{\ast}
=
-\frac{1}{2} \ln z_{\rm c}.
\label{eqn:z2a}
\end{equation}
A result by series expansions is that $z_{\rm c} = 1.09(7)$
for the simple cubic lattice
\cite{Gaunt67}.
It shows that the value of $a^{\ast}$ is negative with
a large error; $a^{\ast} = -0.04(3)$
\cite{Racz80}.

As far as we know,
a Monte Carlo study of a hard-sphere lattice gas
has been published only on the square lattice
\cite{Meirovitch83}.
For the first time we perform Monte Carlo simulations
on the simple cubic lattice.
If we use a Monte Carlo method for
the antiferromagnetic Ising model,
we must simulate the system at very low temperature to estimate
the slope of the critical line at $(T,H) = (0,H_{\rm c})$.
It is difficult to get reliable data.

\section{Monte Carlo simulations}
\label{sec:mcs}
We use the Metropolis Monte Carlo technique
\cite{Binder79,BinderStauffer87} to simulate
the hard-sphere lattice gas (\ref{eqn:gpf})
on the simple cubic lattice
of $V$ sites, where $V$ = $L \times L \times L$
($L$ = 10, 12, 14, 16, 18, 20, 22, and 24),
under fully periodic boundary conditions.

According to Meirovitch \cite{Meirovitch83},
we adopt the grand canonical ensemble.
The algorithm is as follows.
\begin{enumerate}
\item Choose a lattice site.
\label{enum:choose}

\item If its nearest neighbour sites are occupied,
go to \ref{enum:choose}.
If they are vacant, go to the next step.

\item Generate a random number, $r \in [0,1]$.

\item When an atom occupy the lattice site,
we remove it if $r \leq \min (z^{-1},1)$.
When it is vacant,
we add an atom if $r \leq \min (z,1)$.

\item Go to \ref{enum:choose}.
\end{enumerate}

We start each simulation from a large activity with a ground state
configuration and then gradually decrease an activity.
The pseudorandom numbers are generated by the Tausworthe method
\cite{ItoKanada88,ItoKanada90}.
We measure physical quantities at an activity
over $10^{6}$ Monte Carlo steps per site
after discarding $5 \times 10^{4}$ Monte Carlo steps per site
to attain equilibrium.
We have checked that simulations from the ground state configuration
and the configuration without atoms gave consistent results and
there was no hysteresis.
Each run is divided into ten blocks.
Let us the average of a physical quantity, $O$, in each block
$\langle O \rangle_{i}$; $i$ = $1, 2, \ldots , 10$.
The expectation value is
\[
\overline{\langle O \rangle}
=
\frac{1}{10}\,\sum_{i=1}^{10} \langle O \rangle_{i}.
\]
The standard deviation is
\[
{\mit \Delta} \langle O \rangle
=
\left(
\overline{\langle O \rangle^{2}} - \overline{\langle O \rangle}^{2}
\right)^{1/2}/\sqrt{9}.
\]

Let us define a density by
\[
\rho
=
N/V
\]
where $N$ is the number of the atoms
in the lattice of $V$ sites
and an order parameter by
\[
R
=
2\,(N_{\rm A}-N_{\rm B})/V
\]
where $N_{\rm A}$ ($N_{\rm B}$) is the number of the atoms
in the A (B)-sublattice and $N = N_{\rm A} + N_{\rm B}$.
We measure the isothermal compressibility:
\begin{equation}
\kappa
=
V
\overline{
\left(
\langle \rho^{2} \rangle - \langle \rho \rangle^{2}
\right)
/ \langle \rho \rangle
},
\label{eqn:isocom}
\end{equation}
the staggered compressibility:
\begin{equation}
\chi^{\dagger}
=
V
\overline{
\left(
\langle R^{2} \rangle - \langle |R| \rangle^{2}
\right)
}
/ 4,
\label{eqn:stacom}
\end{equation}
and the fourth-order cumulant of $R$ \cite{Binder81}:
\begin{equation}
U
=
1-\frac{1}{3}\,
\overline{
\langle R^{4} \rangle / \langle R^{2} \rangle^{2}
}.
\label{eqn:cum}
\end{equation}

\section{Monte Carlo results}
\label{sec:mcr}
Figure 1 shows the activity dependence of
the isothermal compressibility, $\kappa_{L}(z)$,
defined by (\ref{eqn:isocom}) for various lattice sizes.
The solid curves are obtained by the smoothing procedure of
the {\em B\/}-spline
\cite{Tsuda88}.
As $L$ increases,
the shape of the curve becomes sharper.
In figure 2  we show the activitiy dependence of
the staggered compressibility, $\chi_{L}^{\dagger}(z)$,
defined by (\ref{eqn:stacom}) for various lattice sizes.
It does not seem that the position of the peak shifts
to contrast those of $\kappa_{L}(z)$.
We show the activity dependence of the fourth-order cumulant,
$U_{L}(z)$, of $R$ defined by (\ref{eqn:cum})
for various lattice sizes in figure 3.
There is an intersection between the curves with
the size $L$ and $L+2$.
The positions of these intersections are within a narrow region.

We define effective critical activities,
$z_{\rm max}^{\kappa}(L)$ and $z_{\rm max}^{\chi^{\dagger}}(L)$,
as the peak position of $\kappa_{L}(z)$ and $\chi_{L}^{\dagger}(z)$,
respectively,
and $z_{\rm cross}^{U}(L)$ by
\[
U_{L}(z_{\rm cross}^{U}(L))
=
U_{L+2}(z_{\rm cross}^{U}(L)).
\]
They will converge to the critical activity,
$z_{\rm c}$, as $L \to +\infty$.
We plot them against $1/L$ in figure 4.

We decide to estimate $z_{\rm c}$ from
$z_{\rm max}^{\chi^{\dagger}}(L)$ by the following reasons.
Although $z_{\rm max}^{\kappa}(L)$ seems to
behave systematically for $L$,
it is difficult to extrapolate $z_{\rm c}$ from it
since we need a precise value of a critical exponent $\nu$:
$z_{\rm max}^{\kappa}(L) - z_{\rm c} \sim L^{-1/\nu}$.
We cannot see systematic behaviour for $L$ in
$z_{\rm max}^{\chi^{\dagger}}(L)$ and $z_{\rm cross}^{U}(L)$.
The latter is more scattering than the former.
We get the result, $z_{\rm c} = 1.0588(3)$,
by the arithmetic mean from the data except those of $L$ = 10.
Using (\ref{eqn:z2a}), we have that $a^{\ast} = -0.02857(14)$.
These are consistent with previous results,
$z_{\rm c} = 1.09(7)$
\cite{Gaunt67} and
$a^{\ast} = -0.04(3)$
\cite{Racz80},
within errors.
Our results are more accurate than theirs.

\section{Summary}
\label{sec:sum}
We performed the Monte Carlo simulations of
the hard-sphere lattice gas on the simple cubic lattice.
We estimated the critical activity, $z_{\rm c}$,
to be $1.0588 \pm 0.0003$.
It means that the slope of the critical line
at $(T,H)$ = $(0,H_{\rm c})$
is negative in the antiferromagnetic Ising model
on the simple cubic lattice; $a^{\ast} = -0.02857 \pm 0.00014$.
We conclude that there is no re-entrant phase transition
in this system.

In closing this paper we want to mention that
the antiferromagnetic Ising model is equivalent to
a spin-one Ising model
\cite{KasonoOno92}.
Our result confirms that a re-entrant phase transition occurs
in the latter on the simple cubic lattice.

We have carried out the simulations on two personal computers
with the 486DX2/66MHz CPU and the Linux operating system
(SLS 1.0.3 + JE 0.9.3, Slackware 2.0.0 + JE 0.9.5).

\section*{Acknowledgements}
The author would like to thank Dr. Katsumi Kasono
for useful discussions and critical reading of the manuscript.


\clearpage
\section*{Figure captions}
\begin{description}
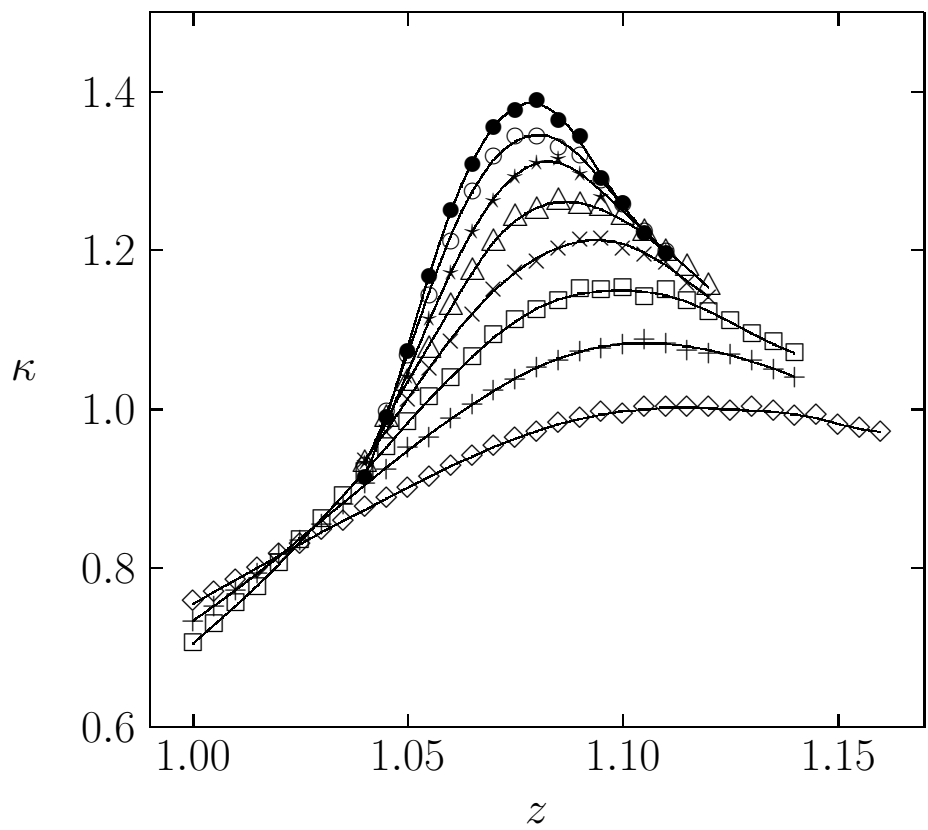
\item[Figure 1] Activity dependence of
the isothermal compressibility,
$\kappa$, defined by (\ref{eqn:isocom})
of the hard-sphere lattice gas on the simple cubic lattice
of $V$ sites under fully periodic boundary conditions.
$V$ = $L \times L \times L$;
$L$ = 10: $\Diamond$, 12: $+$, 14: $\Box$, 16: $\times$,
18: $\triangle$, 20: $\star$, 22: $\circ$, 24: $\bullet$.
The solid curves are obtained by the smoothing procedure
of the fourth-order {\em B\/}-spline.
Errors are less than the symbol size.

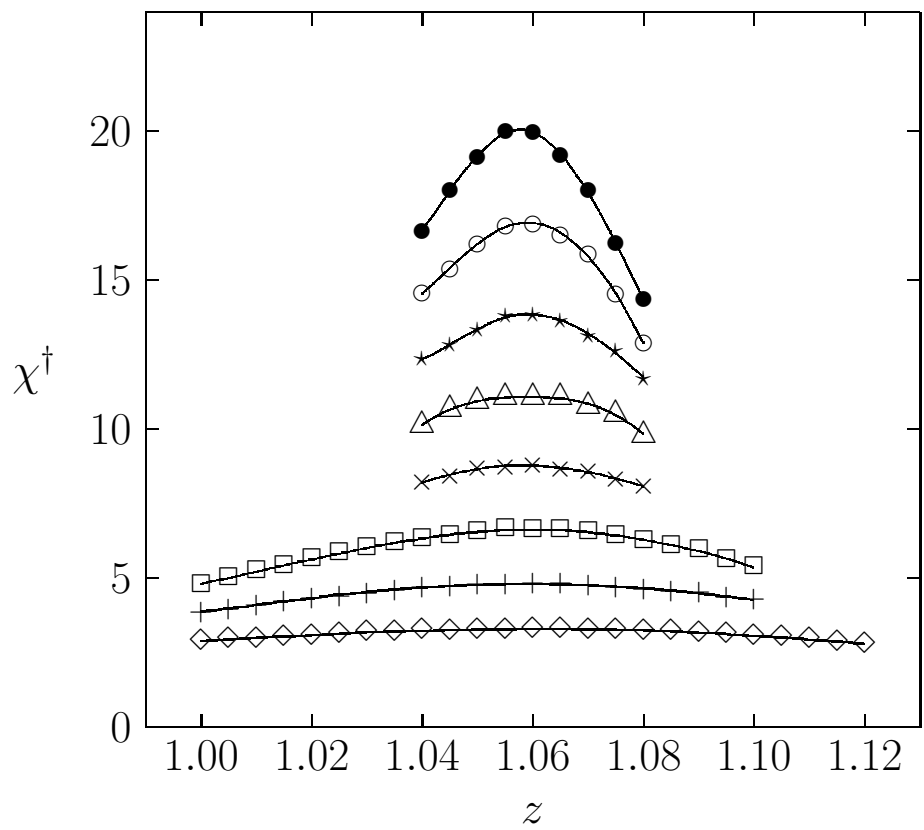
\item[Figure 2] Activity dependence of
the staggered compressibility,
$\chi^{\dagger}$, defined by (\ref{eqn:stacom}).
The meaning of the symbols and the curves is the same as in figure 1.
Errors are less than the symbol size.

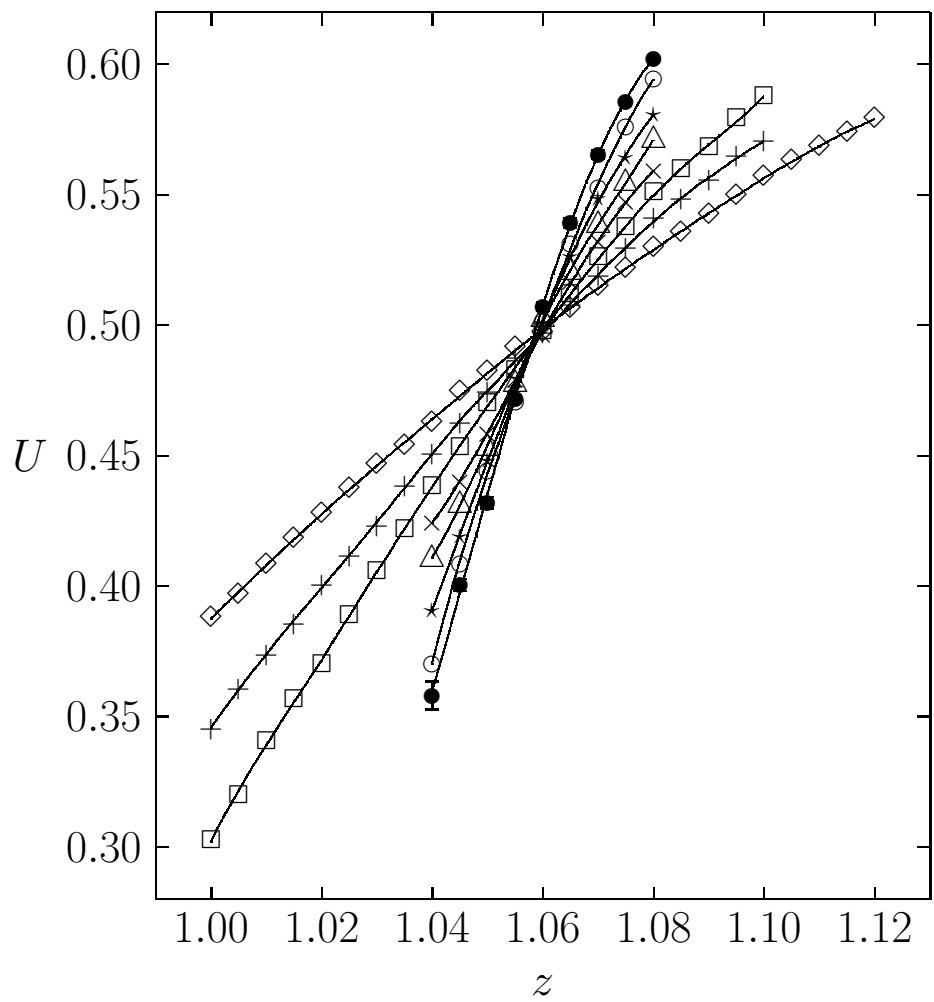
\item[Figure 3] Activitiy dependence of the fourth-order cumulant,
$U$, defined by (\ref{eqn:cum}).
The meaning of the symbols and the curves is the same as in figure~1.

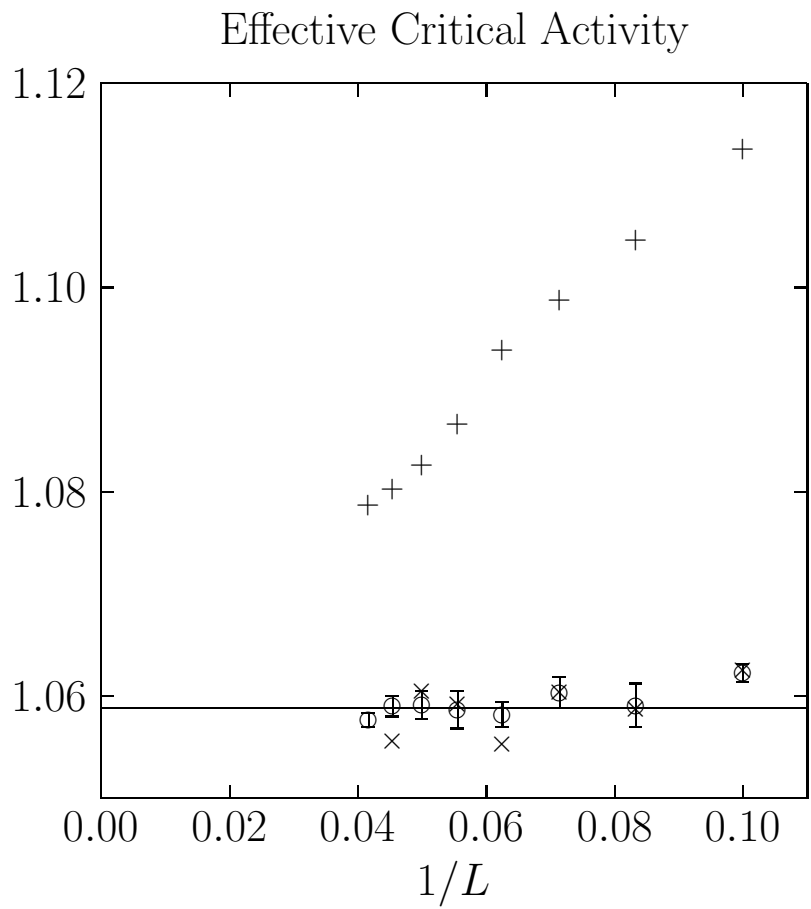
\item[Figure 4] Size dependence of
the effective critical activities,
$z_{\rm max}^{\kappa}(L)$: $+$,
$z_{\rm max}^{\chi^{\dagger}}(L)$: $\circ$,
and $z_{\rm cross}^{U}(L)$: $\times$.
Errors are less than the symbol size
for $z_{\rm max}^{\kappa}(L)$ and $z_{\rm cross}^{U}(L)$.
The horizontal line denotes $z = 1.0588$.
\end{description}

\clearpage
\pagestyle{empty}
\begin{figure}
\begin{center}
\setlength{\unitlength}{0.240900pt}
\ifx\plotpoint\undefined\newsavebox{\plotpoint}\fi
\sbox{\plotpoint}{\rule[-0.200pt]{0.400pt}{0.400pt}}%

\end{center}
\end{figure}
\end{document}